\documentclass[aps,prl,twocolumn,showpacs]{revtex4}
\usepackage{graphicx}
\usepackage{wasysym}

\begin{document}
\def\be{\begin{equation}}
\def\ee{\end{equation}}

\title{Comment on ``Scaling of the linear response in simple aging systems without disorder''}

\author{  Federico Corberi$^\dag$, Eugenio Lippiello$^\ddag$,
and Marco Zannetti$^\S$}
\affiliation{Istituto Nazionale di Fisica della Materia, Unit\`a
di Salerno and Dipartimento di Fisica ``E.R.Caianiello'',
Universit\`a di Salerno,
84081 Baronissi (Salerno), Italy}

\begin{abstract}

We have repeated the simulations of Henkel, Paessens and Pleimling (HPP) [Phys.Rev.E {\bf 69},
056109 (2004)] for the field-cooled susceptibility $\chi_{FC}(t) - \chi_0 \sim t^{-A}$
in the quench of ferromagnetic systems
to and below $T_C$. We show that, contrary to the statement made by HPP,
the exponent $A$ coincides with the exponent $a$ of the linear response function
$R(t,s) \sim s^{-(1+a)}f_R(t/s)$.
We point out what are the assumptions in the argument of HPP that lead them to the 
conclusion $A<a$.

\end{abstract}


\maketitle

\vspace{.2cm}

In a recent paper~\cite{Henkel} Henkel, Paessens and Pleimling (HPP) have addressed the question of the
relationship between the scalings of the linear response function and the zero-field-cooled (ZFC)
susceptibility in ferromagnetic spin systems undergoing aging after a quench to or below $T_C$.
This problem (limited to the quenches below $T_C$) had been previously analysed by us
in a series of papers~\cite{Corberi2001,Corberi2002,Corberi2003,Generic} with the following conclusions:

\begin{itemize}

\item the linear response function

\be
R(t,s)= \left . {\delta \langle \phi (t) \rangle \over \delta h(s)} \right |_{h=0}, \,\,\,  t \geq s
\label{0.1}
\ee
scales as
\be
R(t,s) \sim s^{-(1+a)}f_R(t/s)
\label{1}
\ee
with
\be
a={n \over z} \left ( {d-d_L \over d_U - d_L} \right )
\label{a1.1}
\ee
where $z$ is the dynamical exponent entering the growth law $L(t) \sim t^{1/z}$ of the average defect distance,
$d_L$ is the lower critical dimensionality, $(n=1,d_U=3)$ and $(n=2,d_U=4)$ for scalar or vector order
parameter, respectively.

\item the ZFC susceptibility $\chi_{ZFC}(t,s)  =  \int_s^t ds^{\prime} R(t,s^{\prime})$  scales as
\be
\chi_{ZFC}(t,s) \sim s^{-A}f_{\chi}(t/s)
\label{a2}
\ee
with
\begin{eqnarray}
A=   \left \{ \begin{array}{ll}
        a  \qquad $for$ \qquad   d<d_U  \\
        n/z   \qquad $with log-corrections for$\qquad d=d_U \\
        n/z  \qquad $for$ \qquad d>d_U.
        \end{array}
        \right .
        \label{3}
        \end{eqnarray}

\end{itemize}

\noindent The cases considered by HPP in~\cite{Henkel} correspond to $d<d_U$, where the above result gives
$A = a$~\cite{nota}. Instead, HPP reach the different conclusion $A<a$, thereafter stating that $A$ is a new 
exponent unrelated to $a$ and to aging behavior.
The purpose of the present Comment is to show that the data for the same quench considered in the HPP paper, if properly interpreted,
are in agreement with our finding $A=a$, where $a$ is given by Eq~(\ref{a1.1}).

\begin{figure}
\includegraphics[width=8cm]{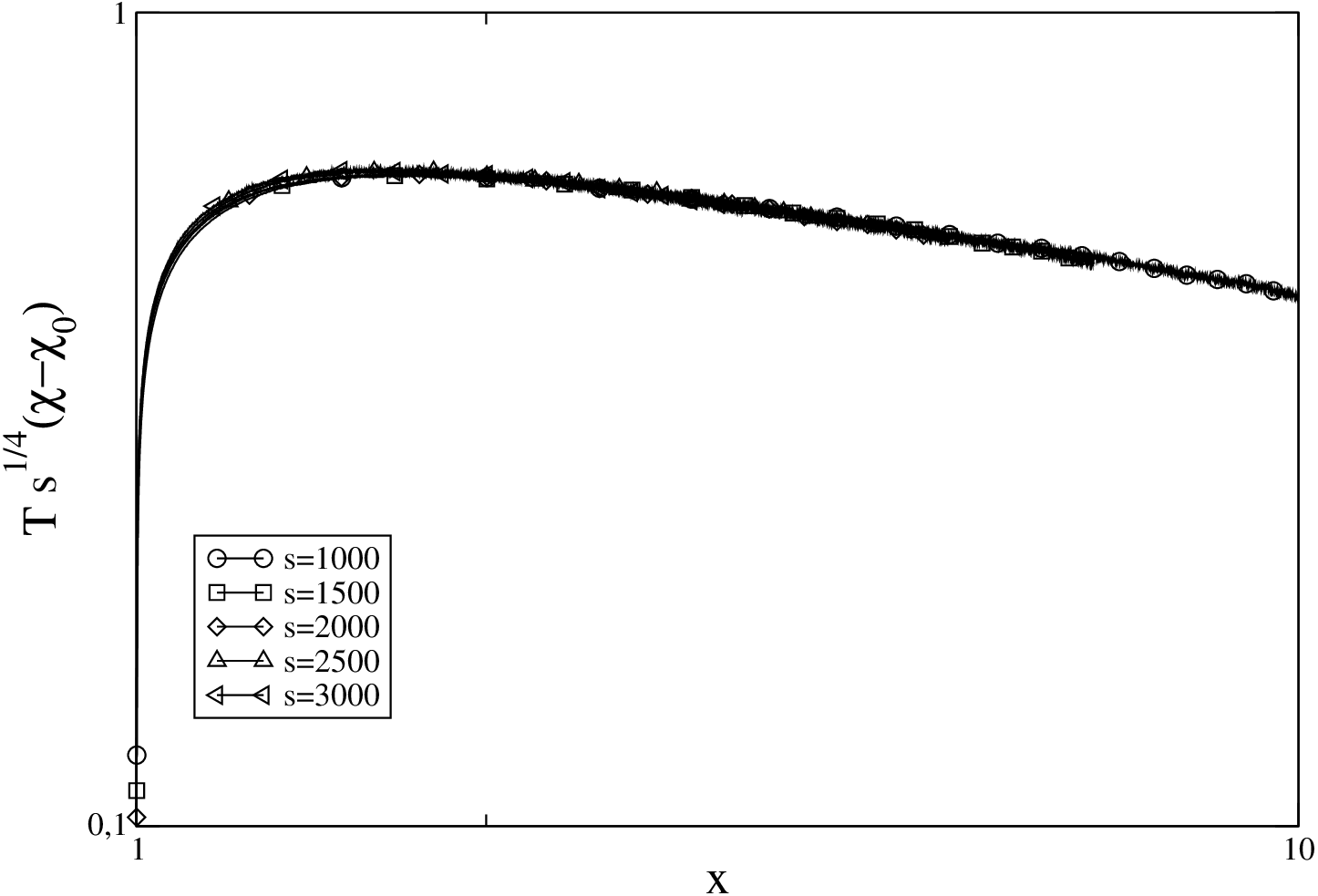}% Here is how to import EPS art
\caption{Data collapse of $s^{1/4}[\chi_{ZFC}(t,s)-\chi_0]$ versus $x$.\\}
\label{fig_a}
\end{figure}

Following HPP, next to the ZFC susceptibility we introduce the field cooled (FC) susceptibility
$\chi_{FC}(t) = \int_0^t ds R(t,s)$ and the thermoremanent (TRM) susceptibility
$\rho_{TRM}(t,s) = \int_0^s ds^{\prime} R(t,s^{\prime})$.
Obviously, the three integrated response functions satisfy the sum rule
\be
\chi_{FC}(t) = \chi_{ZFC}(t,s) + \rho_{TRM}(t,s).
\label{1.4}
\ee
The HHP argument is built on the behavior of the FC susceptibility. From numerical computations they find
\be
\chi_{FC}(t) = \chi_0 + \kappa t^{-A}
\label{2}
\ee
where $\kappa$ is some constant.
Making the distinction between systems of class S, with a finite
equilibrium correlation length $\xi$, and systems of class L, with  $\xi = \infty$, for   $\chi_0$
they make the statement
\begin{eqnarray}
\chi_0 =   \left \{ \begin{array}{ll}
        (1- m_{eq}^2)/T  \qquad $for systems of class S$  \\
        0   \qquad $for systems of class L$
        \end{array}
        \right .
        \label{R1}
        \end{eqnarray}
where $m_{eq}$ is the equilibrium magnetization at the temperature $T$.
Then, assuming that the TRM susceptibility obeys the asymptotic behavior
\be
\rho_{TRM}(t,s) = s^{-a}f_M(t/s)
\label{trm}
\ee
from Eq.~(\ref{1.4}) follows
\be
\chi_{ZFC}(t,s) = \chi_0 + \kappa t^{-A} - s^{-a}f_M(t/s)
\label{b3}
\ee
where for the exponent $a$ they take
\begin{eqnarray}
a =   \left \{ \begin{array}{ll}
        1/z   \qquad $for class S$ \\
        (d-2+\eta)/z  \qquad $for class L$ .        
        \end{array}
        \right .
        \label{4}
        \end{eqnarray}
Let us now look separately to the different cases.

\begin{enumerate}

\item {\it Class S: $2d$ Ising model quenched below $T_C$.}

As an example of class S, HPP consider the $d=2$ Ising model with Glauber dynamics quenched below $T_C$.
Measuring the FC susceptibilty they find that Eq.~(\ref{2}) holds with $\chi_0$ given by
the top line of Eq.~(\ref{R1})
and with $A=1/4$. Then, {\it assuming} that $a$ is given by Eq.~(\ref{4}), that is by $a=1/z=1/2$,
 they make the
statement $A<a$. According to them, $A$ does not have any relationship to aging and 
it is due to the roughness of the interfaces, while $a$ is a subleading exponent. 

Repeating the simulation of HPP with the same quench temperature $T=1.5$, we have reproduced their data for the FC susceptibility, 
However, rather than making an assumption on the value of $a$, we have looked for 
an unbiased comparison of $A$ and $a$, born out of the same set of data. This can be accomplished 
by analysing the ZFC susceptibility according to Eq.~(\ref{b3}).
In Fig.~\ref{fig_a} we have plotted $s^{1/4}[\chi_{ZFC}(t,s)-\chi_0]$ as a function of $x=t/s$ for different values of 
$s$ ranging from $1000$ to $3000$. 
\begin{figure}
\includegraphics[width=8cm]{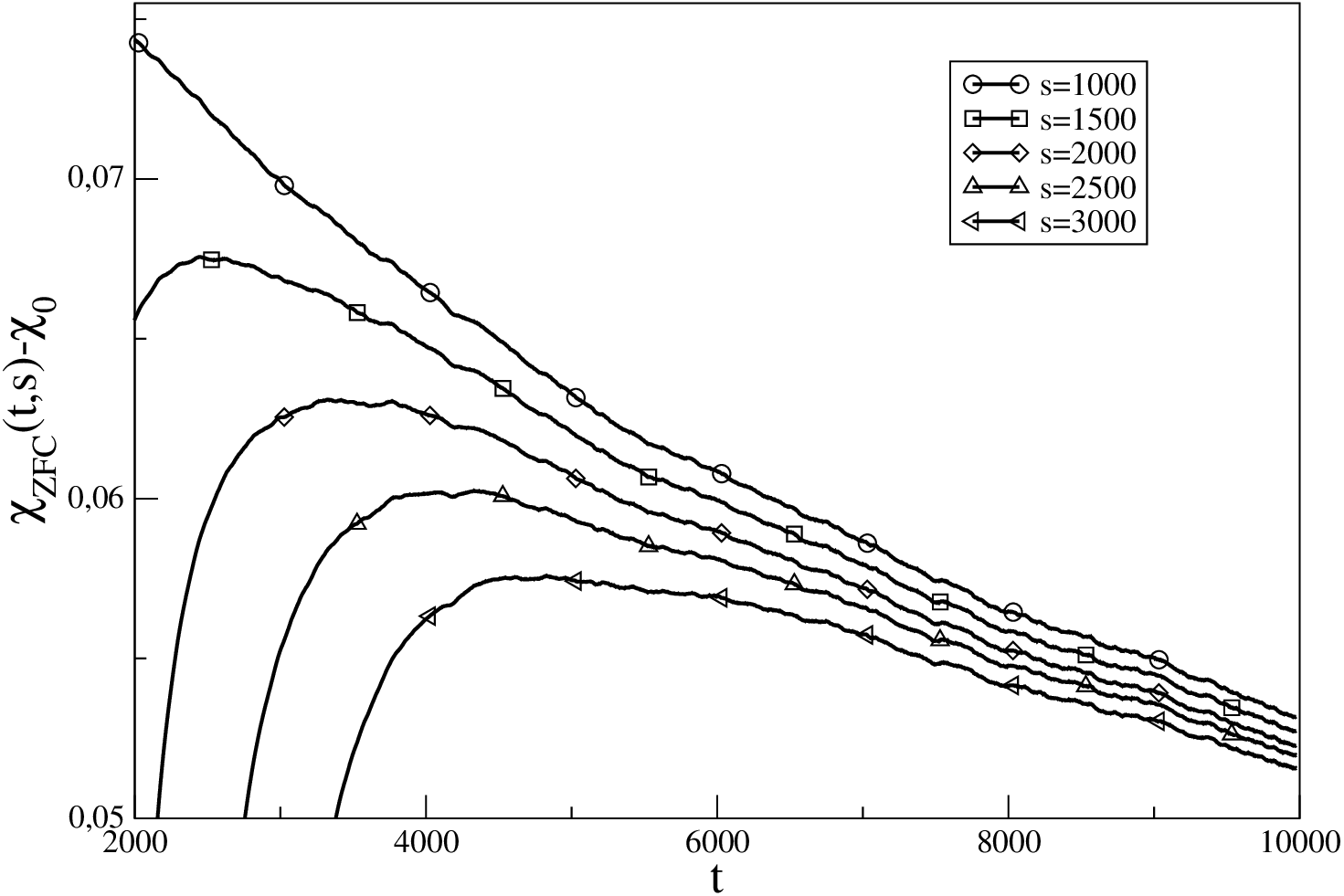}% Here is how to import EPS art
\caption{$\chi_{ZFC}(t,s)-\chi_0$ plotted against $t$.}
\label{fig_b}
\end{figure}
The excellent data collapse in Fig.~\ref{fig_a} has two possible origins: either
$A < a$ and in the range of $s$ considered the third term in the r.h.s. of Eq.~(\ref{b3}) is negligible,
or $a=A$. In order to discriminate between these two possibilities it is enough to replot the same set
of data as a function of $t$ for fixed $s$. If the first alternative is the right one the data should collapse
also in this plot, while in the second case there could be no collapse, due to the existence of the $s$ dependence. 
Fig.~\ref{fig_b} shows that indeed the latter one is the
case, that the $s$ dependence is a large effect and that, therefore, $a=A=1/4$. 

Although the pair of Figures~\ref{fig_a} and~\ref{fig_b} is certainly enough to settle the issue,
in order to help visualize the result we have also produced (Fig.~\ref{fig_c}) the log-log plot of $\chi_{ZFC}(t,s)-\chi_0$
as a function of $s$, for fixed $x$. The data in Fig.~\ref{fig_c} display an excellent fit with a
single power law, with a slope very close to $-1/4$ for every value of $x$,
which makes clear at glance the conclusion  $a=A=1/4$ reached above and in agreement with Eqs.~(\ref{a1.1}) and~(\ref{3}).

In summary, this means i) that $a$ is not subleading as believed by HPP and
ii) that roughening of the interfaces, rather than being unrelated to aging, is 
precisely the mechanism that renders the exponent $a$ smaller (for $d<d_U$) 
than $1/z$, as explained in ref.~\cite{Generic}.  
What goes wrong in the HPP interpretation of the data in~\cite{Henkel} is the
assumption that, for systems of class S, $a$ is given by the top line of Eq.~(\ref{4})~\cite{nota1}.

\begin{figure}
\includegraphics[width=8cm]{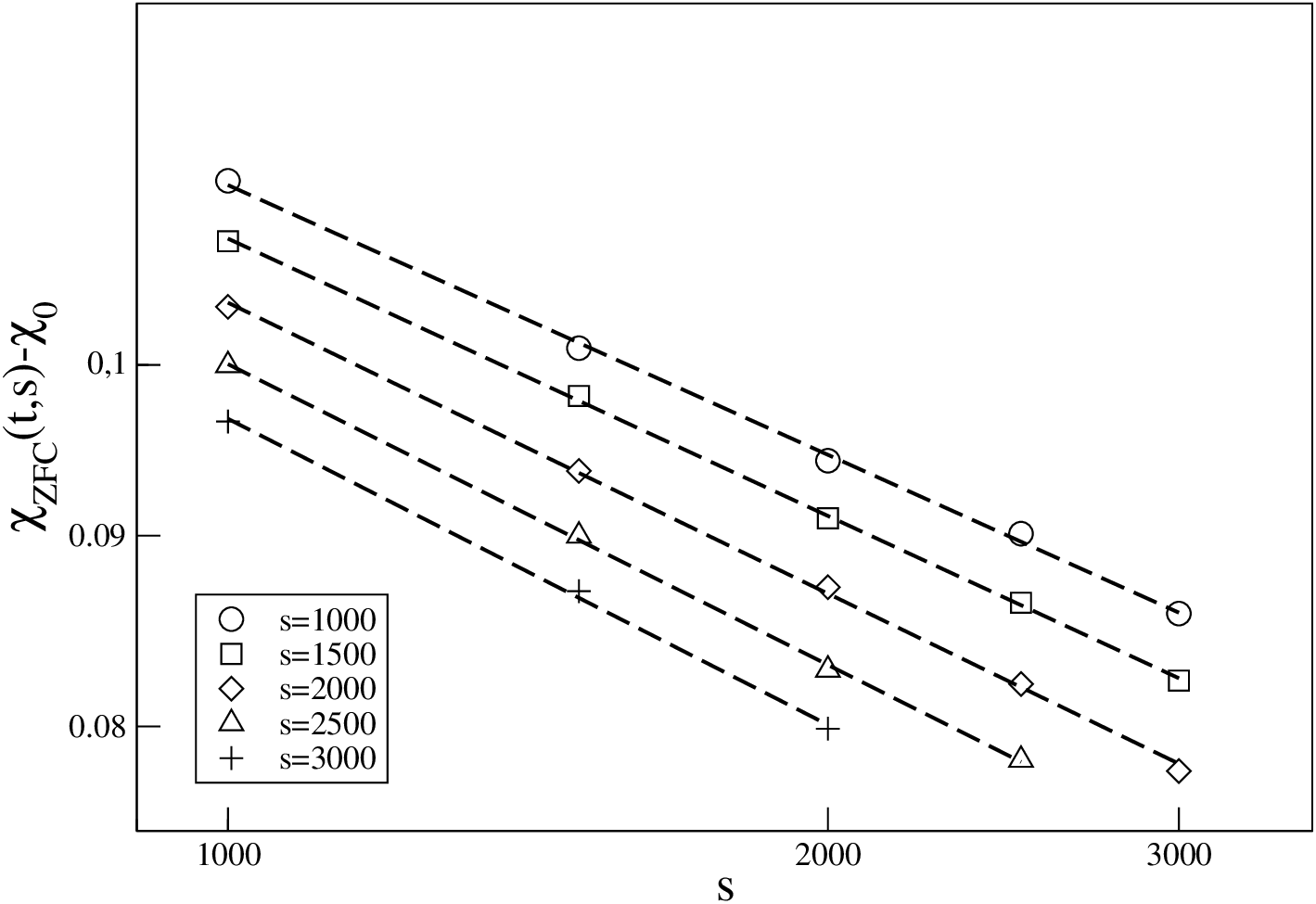}% Here is how to import EPS art
\caption{$\chi_{ZFC}(t,s)-\chi_0$ plotted against $s$ for different fixed values of $x$. 
Straight lines are the best fits (with exponents
0.24,0.25,0.25,0.27,0.27 from top to bottom).}
\label{fig_c}
\end{figure}

\item {\it Class L: $3d$ spherical model quenched to and below $T_C$.}

As an example of class L, HPP consider the spherical model quenched to and below $T_C$.
They compute numerically $\chi_{FC}(t)$ in both cases, with $d=3$, finding
that it saturates to a constant for large $t$. However, they do not identify this constant
with $\chi_0$, since they make the statement~(\ref{R1})
that $\chi_0 =0$ for systems of class L. This is supported by an hand waving argument according
to which $\chi_0$ ought to vanish, since the correlated clusters for systems
of class L should have no ``inside''. Although vague, this assumption
is crucial because it is the starting point of the chain of implications used by HPP:
the saturation of $\chi_{FC}(t)$ to a constant value {\it and} $\chi_0 =0$ imply $A=0$, which in turn implies $A<a$,
since in the spherical model $a=(d-2)/2$, both in the quench to and below $T_C$.

Before going further, it is necessary to clarify the physical meaning of the constant $\chi_0$
appearing in Eq.~(\ref{2}).
This can be readily understood recalling that
\be
\lim_{t \to \infty}  \chi_{FC}(t) = \chi_{eq} = (1-m_{eq}^2)/T
\label{b5}
\ee
where $\chi_{eq}$ is the static susceptibility. Since the second equality in the above equation is nothing but the static
fluctuation-dissipation theorem, it holds for all systems, be they of class S or class L,
quenched below $T_C$ or to $T_C$.
Therefore, from Eqs.~(\ref{2}) and~(\ref{b5}), we can identify $\chi_0=\chi_{eq}$
and Eq.~(\ref{R1}) must be replaced by
\be
\chi_0 =  (1- m_{eq}^2)/T
\label{fdt}
\ee
both for systems of class S and class L, which clearly reduces to $\chi_0 = 1/T_C$ for the quenches to $T_C$.
Therefore, $\chi_0=0$ assumed by HPP is excluded in all cases.

Furthermore, in the case of the spherical model there is no room for assumptions, since the model
is exactly soluble. Using the formulas that HPP give in~\cite{Henkel}, it is not difficult to
derive analytically the large $t$ behavior of the FC susceptibility obtaining
\be
\chi_{FC}(t) = (1-m_{eq}^2)/T + \kappa t^{-(d-2)/2}
\label{FC}
\ee
with $m_{eq}^2 = 1-T/T_C$, $\kappa= [\Gamma(1-d/2)\Gamma(1+\omega/2)] / \Gamma(2 - {d-\omega \over 2})$
where $\Gamma$ is the gamma function, $\omega=d/2-2$ in the quench to $T_C$ and $\omega=-d/2$ in the quench below $T_C$.
Therefore, comparing with Eq.~(\ref{2}), we have $\chi_0 =\chi_{eq} = (1-m_{eq}^2)/T$~\cite{nota2}, as expected from Eq.~(\ref{fdt}).
This implies $A=a=(d-2)/2$ for $T \leq T_C$, in agreement with Eqs.~(\ref{a1.1}) and~(\ref{3}).

\item {\it Class L: $2d$ Ising model quenched to $T_C$.}

As an additional instance of a system of class L, HPP consider the  $2d$ Ising model quenched to $T_C$.
Again, they find that the FC susceptibility saturates to a constant. By the same reasoning as in the
previous case, from the assumption that $\chi_0$ ought to vanishes they make to descend  $A=0 < a=(d-2+\eta)/z_c=0.115$,
where we have used $z_c=2.167$~\cite{z_c} and the exact result $\eta=1/4$.

Although the argument of Eq.~(\ref{b5}) ought to suffice,
we have repeated their simulations and we have plotted (Fig.~\ref{fig_d}) $\log (\chi_0-
\chi_{FC}(t))$ against $\log t$,
with $\chi_0=1/T_C$.
The figure shows a very clean power law decay with $A=0.115 \pm 0.005$, which compares very well with the value of $a$
given above and yields, again, $A=a$.
Furthermore, the observation of the decay with the correct value of the exponent implies that also the subtraction by
$\chi_0=1/T_C$ is the correct one.

\end{enumerate}

In summary, the unbiased analysis of the data for the FC susceptibility in all cases considered by HPP,
that is in systems of class S and class L with $d<d_U$, yields $A=a$ in agreement with Eqs.~(\ref{a1.1}) and~(\ref{3}).
The biases in the HPP analysis, which lead to the wrong conclusion $A<a$, are in the two assumptions
i) $a=1/z$ for systems of class S and ii) $\chi_0=0$ for systems of class L.

\vspace{5mm}

{\bf Acknowledgments}

This work has been partially supported by MURST through PRIN-2002. \\

\begin{figure}
\includegraphics[width=8cm]{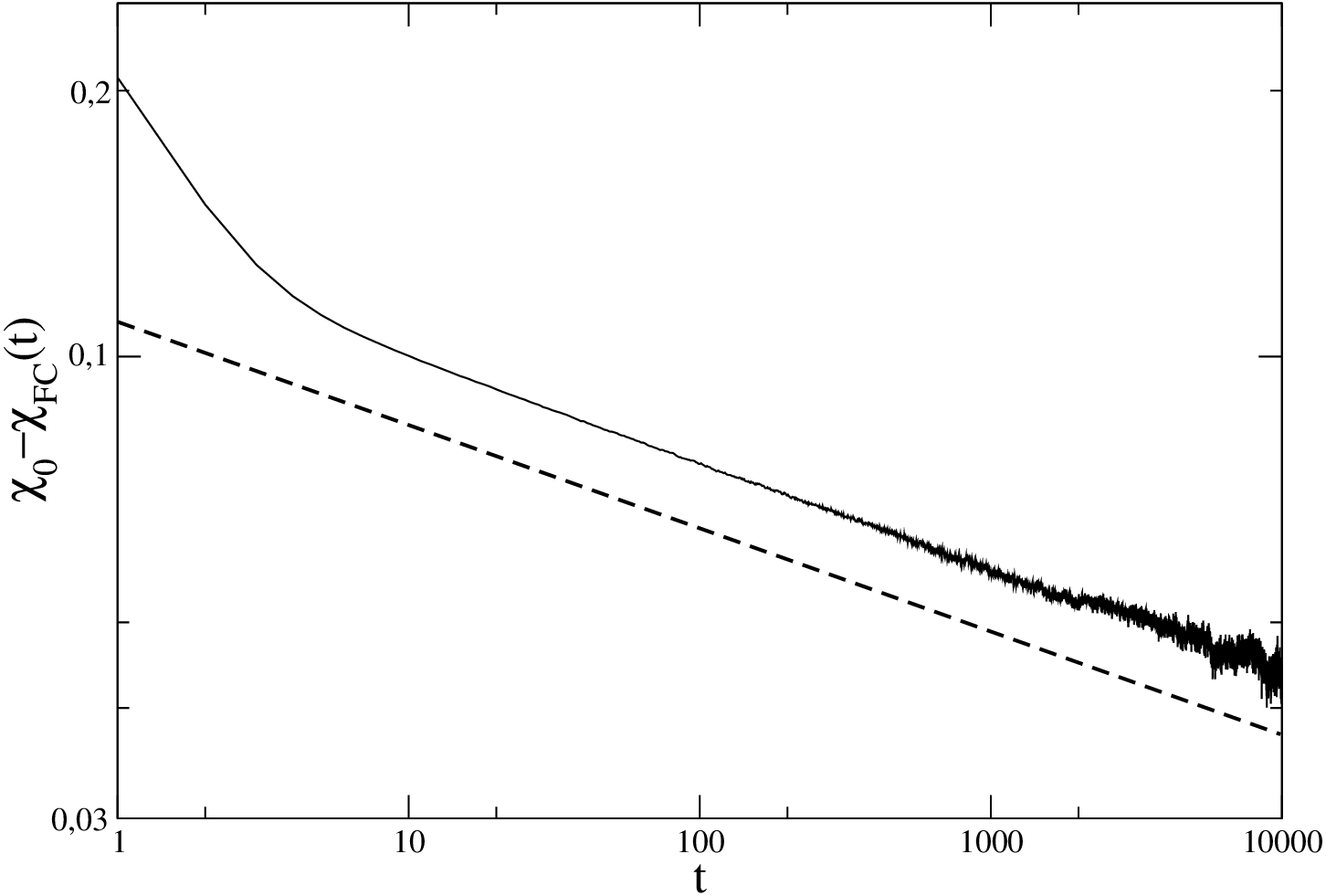}% Here is how to import EPS art
\caption{The field cooled susceptibility $\chi_{FC}(t)-\chi_0$ vs $t$ for the 
2D Ising model at $T=T_c \simeq 2.269$. The straight line is the expected behavior $\chi_{FC}(t)-\chi_0 \sim t^{\eta/z_c} $ with $\eta =1/4$ and $z_c=2.167$.} 
\label{fig_d}
\end{figure}

\dag corberi@sa.infn.it \\
\ddag lippiello@sa.infn.it \\
\S zannetti@na.infn.it \\


\begin{thebibliography}{99}
\bibitem{nota2}
Notice that, in the particular case of the spherical model, this gives $\chi_0=1/T_C$ for $T \leq T_C$, since
thermal fluctuations are critical for all $T \leq T_C$.
\end{thebibliography}
\end{document}